\documentclass{article}
\usepackage[utf8]{inputenc}
\usepackage[a4paper,top=2.5cm,bottom=2cm,left=2cm,right=2cm,marginparwidth=1.75cm]{geometry}
\usepackage{tabularx}
\usepackage{amsmath}
\usepackage[colorinlistoftodos]{todonotes}
\usepackage[colorlinks=true, allcolors=purple]{hyperref}
\usepackage{adjustbox}
\usepackage{xcolor}
\usepackage{caption} 
\usepackage{verbatim}
\usepackage{booktabs}
\usepackage{chronology}
\usepackage{tikz}
\usepackage{natbib}
\usepackage{ragged2e}
\usetikzlibrary{snakes}
\usepackage[doublespacing]{setspace}
\usepackage{lscape}
\usepackage{subcaption}
\usepackage{footnote}
\usepackage{longtable}
\usepackage{graphicx}
\usepackage{breqn}
\graphicspath{{CoefPlots/}}
\newcolumntype{C}[1]{>{\centering\arraybackslash}p{#1}}
\usepackage{multirow}
\usepackage[title]{appendix}
\makeatletter
\newcommand{\ssymbol}[1]{^{\@fnsymbol{#1}}}
\makeatother
\usepackage{float}
\usepackage[flushleft]{threeparttable}
\usepackage{makecell}
\usepackage{chngcntr}
\counterwithin{table}{section}
\usepackage{url}

\floatstyle{plaintop}

\restylefloat{table}
\restylefloat{figure}
\title{Impact of the Availability of Chat-GPT on Software Development Activity}
\author{Alexander Quispe\thanks{The World Bank} \and Rodrigo Grijalba\thanks{Pontificia Universidad Católica del Perú}} 
\date{}

\begin{document}
\maketitle

\begin{center} 
First version: March 1, 2024 

This version: \today 
\end{center} 

\begin{abstract}
    Advancements in Artificial Intelligence, particularly with ChatGPT, have significantly impacted software development. Using novel data from GitHub Innovation Graph, we hypothesize that ChatGPT improves software production efficiency. Utilizing natural experiments where some governments banned ChatGPT, we employ Difference-in-Differences (DID), Synthetic Control (SC), and Synthetic Difference-in-Differences (SDID) methods to estimate its effects. Our findings indicate a significant positive impact on the number of git pushes, repositories, and unique developers per 100,000 people, particularly for high-level, general-purpose, and shell scripting languages. These results suggest that AI tools like ChatGPT can substantially boost developer productivity, although further analysis is needed to address potential downsides such as low-quality code and privacy concerns.
\end{abstract}


\clearpage 

\section{Introduction}

This past decade, a significant proportion of technological progress in general has been marked by accelerating progress in Artificial Intelligence. As reported by the World Intellectual Property Organization (2022) \cite{WIPO:2022}, the growth rate of Artificial Intelligence patents has been eight times that of all patents, signaling this field's significantly increased interest and potential. Language Processing and Generation is an important sub-field that creates algorithms capable of analyzing, predicting, and generating text by training Machine Learning models with human language. Many approaches have been used for this end. Still, the most capable models currently use the transformer architecture, which can train and be deployed quickly by focusing on relationships between words in natural language. 

One of the most important examples of Large Language Models (LLMs) developed through this approach has been the GPT family of models. The developing company, OpenAI, has the stated goal of advancing AI towards Artificial General Intelligence (AGI) in favor of furthering socially beneficial technological development (OpenAI, 2018) \cite{OpenaAI:2018}. As part of this goal, they have developed one of the most advanced and capable family of models for language processing, which is also used for text generation in their chatbot application, ChatGPT. Thanks to the flexibility of the model and the fact that it has been trained in human language from a large database containing diverse topics, including programming questions, the chatbot has been used to aid in software development. 

Given the popularity of the application and its capabilities for software development, it would be expected that the process of creating software would be made noticeably more efficient, which would be reflected in an increase in the production of software, version, patches, etc. after the application has been made available. Such a measurement would add to the emerging literature on using ChatGPT to aid in software development. To make this measurement, we can exploit the fact that, initially, ChatGPT was made available only in a subset of countries, which can be considered a treatment group; the remaining groups can serve as a control group. With the availability of data on software development at the country level, we have the necessary pieces to analyze the availability of ChatGPT as a treatment in a natural experiment. We use the Difference in Differences, Synthetic Control, and Synthetic Difference in Differences approaches to analyzing quasi-experimental data to estimate ChatGPT's effect on software production in each country-quarter. We find a significant and essential change in the number of pushes per capita in countries with ChatGPT availability compared to those without it; this suggests that the speed of software production is increased by ChatGPT's ability to generate suggestions for programming algorithms, designing software architecture, and repairing bugs, among other uses in software development.

The rest of this article is organized as follows. Section 2 reviews the literature on the potential, current usability, and measured impact of ChatGPT as a tool for aiding in the software development process. Section 3 first describes the country-level data on software development activity that will be used, namely, GitHub's Innovation Graph. Then, it explains the methods through which the data will be used to estimate the effect ChatGPT has had, on average, on each country's production of software: Difference in Differences, Synthetic Control, and Synthetic Difference in Differences. Section 4 shows the results of each method on the variables we construct using the available data. Finally, section 5 discusses these findings in the current context of rapid progress in AI.
\section{Literature review}

The use of machine learning methods, specifically neural networks, to aid productivity in software development is a very rare practice. The use of ChatGPT for these purposes is even more recent, with its launch in November 2022 OpenAI, 2022 \cite{OpenAI:2022}. As is expected for very recent phenomena, the literature on the impact of ChatGPT on software development productivity is very scarce. Nonetheless, the undeniable disruptiveness of the application has awakened great interest in its study, which has offset some of the scarcity in the literature. Still, research on the potential, capabilities, and impact of ChatGPT on software development is still in its infancy, and the constantly evolving nature of the topic requires quick adaptations of research to update the understanding of present and future questions adequately. The relevance of this last point is illustrated by the fact that the vast majority of the literature on this topic pertains to the GPT-3.5 model, even though GPT-4 was launched in March 2023.

\subsection{Potential of ChatGPT on Software Development}

Abu Jaber, Beganovic, and Abd Almisreb (2023) \cite{AbuJaber:2023} conduct a literature review on the potential of ChatGPT in software development. They outline potential uses for troubleshooting and bug repair and educational applications for these purposes. This includes optimizing programs and numerical algorithms, as well as the potential incorporation of ChatGPT as an integrated controller in Generalized Intelligence. The authors discuss the creation of software solutions and architectures, both comparatively (evaluating multiple responses) and through a dialogical process with the application for software architectural design. They also delve into considerations for Prompt Engineering when employing the program for software development.

Rahmaniar (2023) \cite{Rahmaniar:2023} comes to similar conclusions. The author surveys the potential for productivity gains in software development through assistance in programming, documentation creation, training and onboarding, code review, and interactions with clients and stakeholders. All this thanks to the model's ability to process natural language, generative capacity, flexibility and potential for model learning, dialogical interactivity, the large spectrum of applicability for its use, and contributions to the Open Source space. The author also discusses the negative ramifications of such uses, such as possible incompleteness of generated code, which may cause difficulties for users and potential security vulnerabilities. In this regard, there is also the potential for malicious use of the model to create malware. The author calls for considering ethical considerations; for example, during the learning process, models may be exposed to content that integrates and perpetuates biases with negative consequences. With future improvements in language models, especially with the potential of GPT models, response generation would have better performance, translating into greater ease and productivity gains for developers using LLM tools. Future versions could be used to streamline the development process, where generation is constant to optimize and repair code in real-time.

\subsection{Tests on ChatGPT's capabilities:}

One of the potentials of ChatGPT in software development lies in its usefulness in education.  Jalil et al. (2023)\cite{Jalil:2023} test ChatGPT with solved problems from Ammann and Offutt (2016), a widely used introductory textbook on software testing. They find that the model can respond correctly to about 43\% of software testing questions and provides a sufficient explanation for its response in most cases.

Another possible avenue for innovation and productivity gains is using ChatGPT to optimize algorithms and numerical methods. Badini et al. (2023) \cite{Badini:2023} use ChatGPT for troubleshooting and optimization of GCode programs for additive manufacturing (3D printing) routines. After 6 iterations of dialogue with the chatbot with the purpose of training on the low-level GCode language, they prompt ChatGPT to solve manufacturing problems resulting from existing routines, and find noticeable improvements in the build quality of the end products after using ChatGPT's output.

As for numerical methods, Kashefi and Mukerji (2023) \cite{Kashefi:2023} propose and test the usage of ChatGPT on the generation of code for several numerical methods related to physics, in several programming languages. This involves programming the simulation of simple linear systems as well as more complicated dynamic systems that involves one or more differencial equations. ChatGPT was able to generate sufficient, programs for the mathematical system described by the researchers. Some issues arise like the generation of singular matrices, attempts at opperating on incompatible arrays, and interruptions of the generation of code when the program is lengthy.

It is also possible to use ChatGPT as a controller for routing tasks and sub-tasks to algorithms more suitable for solving them. This is proposed and tested by Shen et al. (2023) \cite{Shen:2023}, who use GPT-4 in conjunction with several specialized models from the HuggingFace platform to solve complex tasks that involve the solution of several specialized sub-tasks. They develop a 4 stage architecture called HuggingGPT: first, ChatGPT generates a plan for solving the complex task through smaller sub-tasks; on the second stage, it chooses the most capable HuggingFace model to solve each sub-task; in stage 3, the sub-tasks are routed to their corresponding models and these, in turn, generate the corresponding output; the fourth and final stage involves parsing these outputs and presenting them to the user. They test this procedure on image captioning, image classification, and object detection. They find that GPT-4 is highly accurate at creating the necessary task plans and classifying sub-tasks to the corresponding specialized models. They also successfuly test HuggingGPT on a variety of complex tasks involving text, images, audio and video.

Ahmad et al. (2023) \cite{Ahmad:2023} describe and test human-bot collaboration for the purpose of formally outlining sofware architecture. Their testing starts with outlining an architecture story that describes the conditions that must be satisfied by the software. This outline is fed to the chatbot as an initial prompt as a preliminary step for the testing. After this, a novice software architect enters a dialectical process with the chatbot to analyze, synthesize, and evaluate a potential architectural solution. They conclude that this dialectica, collaborative process can be easily carried out to outline a software architecture more efficiently in the early stages of product development.

Finally, Sobania et al. (2023) \cite{Sobania:2023} and Zhang et al. (2023) \cite{Zhang:2023} test the bug repair capabilities of ChatGPT using the QuickBugs database and a custom database, respectively. Both articles find overall success using the model when prompting it to fix the bugs in the code presented, and that more rounds of dialogue increase the success rate. Sobania et al. (2023) specifically find that ChatGPT is comparable to other LLMs specialized in this task, and that LLMs outperform a standard, RN-based program repair application. On the other hand, Zhang et al. (2023) create a new database of coding problems to test the generalization capabilities of the model, as QuickBugs may have been used for training GPT-3.5. They again find that ChatGPT is capable of giving correct fixes in most scenarios.

\subsection{The Impact of ChatGPT on Software Development Productivity}

Gallea (2023) \cite{Gallea:2023} compares the Stack Overflow pages for Python and R, both programming languages used in data science. The reason behind this comparison is the fact that Python is a much more popular programming language with a high amount of potential training material for the model, and the author mentions the use of ChatGPT was not efficient when answer R related question; this would imply a lower impact of ChatGPT on the R language. The data comes from Stack Overflow Explorer, specifically information on the number of questions, the resolution status of the questions, and the score for each of them. The results from the Differences in Differences analysis suggest a negative impact on the quantity of questions, a positive impact on the average score (as a measure of quality) of questions, and a negative impact on the proportion of resolved questions (as a measure of complexity) for Python compared to R.

Saguu and Ante (2023) \cite{Saggu:2023} analyze the returns of crypto-assets related to artificial intelligence. For this purpose, they utilize daily frequency data of crypto-asset prices from Coingecko and CoinMarketCap. The treatment group is considered as those assets related to artificial intelligence, and the start of the post-treatment period is the launch of ChatGPT. They also control for total market capitalization and transaction volume for each asset. They employ both the Differences in Differences and Synthetic Differences in Differences methods. They find that the launch of ChatGPT had a positive impact on the return of those assets related to artificial intelligence on both platforms, compared to those unrelated to AI.

Demirci, Hannane and Xinrong (2023) \cite{Demicri:2023} analyze the impact of ChatGPT on the demand for services for freelance workers on a platform dedicated to facilitating such services. Firstly, they identify service groups based on the skills required to carry them out. Next, they assign an artificial intelligence exposure index to each type of service, indicating the feasibility of using AI to perform services in each group. The analysis is further adjusted using the Google Search Volume Index. They apply the Differences in Differences method between two groups: high and low exposure to artificial intelligence. They find that the impact of ChatGPT was negative on the quantity of postings seeking services in groups with high exposure to AI, compared to those with a low exposure index.

Del Rio-Chanona, Laurentsyeva and Wachs (2023) \cite{Delriochanona:2023} present another example of an analysis of the impact of ChatGPT on question and answer platforms for programming topics. In this case, the authors compare Stack Overflow with Math Exchange, a platform focused on questions and answers about mathematics; the topics on this platform would be less susceptible to the effects of ChatGPT. They also compare Stack Overflow with its Russian counterpart; and with Segmentdefault, its Chinese counterpart. This is because access to ChatGPT is restricted in these two countries, which would decrease the impact it can have on the software developmnet field. By applying Differences in Differences, the authors find a negative impact on the number of posts per week, the number of questions per week, and the number of posts on weekdays on Stack Overflow compared to the other platforms.

Finally, Kreitmeir and Raschky (2023) \cite{Kreitmeir:2023} analyze the impact on productivity in software development using daily user-level data provided by GitHub. They leverage the ban of ChatGPT in Italy, comparing it to France and Austria, where the application is freely accessible. As productivity measures, the authors use the existence of new releases by each user; the sum of pushes, pull requests, comments on pull requests, comments on commits, repo creations, and issues; the sum of pushes and pulls; and the total number of events by each user. They also analyze the quantity of new Tor users, as users in countries where access is restricted may bypass the ban through this means, indicating high demand and recognition of the application's utility. The result obtained through the Differences in Differences method indicates a significant negative impact on the probability that each user-day presents a new software release in Italy compared to other countries. This suggests lower productivity as a result of the prohibition.

\section{Methods}

\subsection{Data}

The entirety of the data come from \href{https://github.com/github/innovationgraph}{GitHub's Innovation Graph}, specifically the table that contains the number of pushes per country and per quarter. The data spans form the first quarter of 2020 until the first quarter of 2023. The information is also separated by jurisdiction, including a total of 179 units. Of these, we keep 151 after dropping 28 units that lack observations for the whole time span. We also remove the observations related the EU as an aggregate, as they cause perfectly collinear with those of the individual EU countries. Finally, we also exclude Hong Kong from our analysis, because it shows an atypically high level of pushes compared to all other countries. We consider the start of the treatment period to be the fourth quarter of 2022, as the initial launch of the app came out on November 30th of the same year. The treated units are the countries that were originally announced to have access to the application, and during the time span included, no new countries were added to the treatment group. With this, we have a case of block design as described by Arkhangelsky et al. (2021) \cite{Arkhangelsky:2021}, which facilitates the computation of the estimators. This design consists of 120 treatment units and 27 control units; 11 pre-treatment periods and 2 treatment periods. One result shown by Abadie, Diamond and Hainmueller (2010) \cite{Abadie:2010} for the Synthetic Control estimator -- which also holds for the SDiD estimator -- is that the upper bound of the magnitude of the estimation bias for the Average Treatment Effect grows with the number of control units and shrinks with the number of pre-treatment periods. We assess that the data presented should render a relatively unbiased estimator.

\subsection{Estimation and inference}

We may represent the data for each variable as a $N\times T$ matrix, with $N$ units observed for $T$ time periods. We have $N_0$ control units and $N_1 = N-N_0$ treatment units. Analogously, we have $T_0$ pre-treatment periods and $T_1=T-T_0$ post-treatment periods. The models we use can be generally described as specific cases of the following equation:

\begin{equation}
\widehat{ATE} = \frac{1}{N_1}\sum_{i=N_0+1}^N\left(\frac{1}{T_1}\sum_{t=T_0+1}^TY_{it}-\frac{1}{T_0}\sum_{t=1}^{T_0}\lambda_iY_{it}\right) - \sum_{i=1}^{N_0}\omega_i\left(\frac{1}{T_1}\sum_{t=T_0+1}^TY_{it}-\sum_{t=1}^{T_0}\lambda_tY_{it}\right)
\end{equation}

Where $Y_{it}$ is the outcome variable for unit $i$ at time period $t$; $\omega_i$ is the factor loading for unit $i$, such that $\sum_{i=1}^{N_0}\omega_i=1$; and $\lambda_t$ is the factor loading for time period $t$, such that $\sum_{t=1}^{T_0}\lambda_t=1$. The first term in the right side is the difference between the average outcome level before and after the treatment for the treatment group. The second term of the right side is this same measure but for the control group. What differentiates each model is the way the unit and time weights are calculated.

\subsubsection{Difference in Differences}

Neither the units nor the time periods are weighted, which is equivalent to their weights being defined by:

\[
\omega_i=\frac{1}{N_0};\forall i=1,...,N_0
\]
\[
\lambda_t=\frac{1}{T_0};\forall t=1,...,T_0
\]

\subsubsection{Synthetic Control}

The vector $(\omega_1, \ldots, \omega_{N_0})$ is defined as:
\[
(\omega_1, \ldots, \omega_{N_0}) = \arg\min_{\omega_1, \ldots, \omega_{N_0}} \sum_{t=1}^{T_0} \left( \sum_{i=1}^{N_0} \omega_i Y_{it} - \frac{1}{N_1} \sum_{j=1}^{N_1} Y_{jt} \right)^2
\]

Subject to:

\[\sum_{i=1}^{N_0}\omega_i=1;\omega_i\geq0;\forall i=1,...,N_0\]

On the other hand, SC only takes into account the unweighted treatment period averages, and not the difference between these and the pre-treatment period. This means that $\lambda_t=0$ for all pre-treatment periods.

\subsubsection{Synthetic Difference in Differences}

In this case, we take the vector $\omega_0,\omega_1,...,\omega_N$, which includes factor loadings for the treatment units and $\omega_0$, an intercept term. They are defined by:

\[\omega_i=\frac{1}{N_1};\forall i=N_0+1,...,N\]

\[
(\omega_0, \ldots, \omega_{N_0}) = \arg\min_{\omega_0, \ldots, \omega_{N_0}} \left( \sum_{t=1}^{T_0} \left( \omega_0 + \sum_{i=1}^{N_0} \omega_i Y_{it} - \frac{1}{N_1} \sum_{i=N_0+1}^{N} Y_{it} \right)^2 + \zeta^2 T_0 \left\| (\omega_0, \ldots, \omega_{N}) \right\|_2^2 \right)
\]

Subject to:

\[\sum_{i=1}^{N_0}\omega_i=1;\omega_i\geq0;\forall i=1,...,N_0\]

The rightmost term is for regularization, where:

\[
\zeta=(N_1T_1)^{1/4}\hat{\sigma}
\]
\[
\hat{\sigma}^2=\frac{1}{N_0(T_0-1)}\sum_{i=N_0}^{N_0}\sum_{t=1}^{T_0-1}(\Delta_{it}-\overline{\Delta})^2
\]
\[\Delta_{it}=Y_{i(t+1)}-Y_{it}\]
\[
\overline{\Delta}=\frac{1}{N_0(T_0-1)}\sum_{i=1}^{N_0}\sum_{t=1}^{T_0-1}\Delta_{it}
\]

Arkhangelsky et al. (2021) explain how this regularization factor is useful to account for unit-level, time-correlated outcomes. On the other hand across-unit, time-level correlation is considered to be captured by systematic variations in the outcome. Therefor, regularization is not included for the estimation of $\lambda_i$. In the case of this model, the vector $(\lambda_0,...,\lambda_{T_0})$ is defined as follows: 

\[
(\lambda_0, \ldots, \lambda_{T_0}) = \arg\min_{\lambda_0, \ldots, \lambda_{T_0}} \sum_{i=1}^{N_0} \left( \lambda_0 + \sum_{t=1}^{T_0} \lambda_t Y_{it} - \frac{1}{T_1} \sum_{t=T_0+1}^{T} Y_{it} \right)^2
\]

Subject to:

\[\sum_{i=1}^{T_0}\lambda_t=1;\lambda_t\geq0;\forall t=1,...,N_0\]

\subsubsection{Inference}
The data we use is unique in that we have the advantage of including a big number of treated units, with exposition happening simultaneously in all of them. This gives us the opportunity of using the bootstrap estimator, which was demonstrated by Arkhangelsky et al. (2021) to be asymptotically normal with a large number of treated units. We refer to Clarke et al. (2023) \cite{Clarke:2023} for the design of the bootstrap algorithm to estimate the standard error of the ATE estimators.

\section{Results}

The table presented in \ref{tab:gpt_push_repo_dev} provides an analysis of the impact of ChatGPT on software development activities, measured by three key outcomes: the number of git pushes, the number of repositories, and the number of unique developers, each standardized per 100,000 people in the population. This analysis employs three methodological approaches: Difference-in-Differences (DID), Synthetic Control (SC), and Synthetic Difference-in-Differences (SDID). The coefficients reported for "Chat GPT Available" represent the estimated effect of ChatGPT's availability on these outcomes.

The results in Panel A indicate a significant positive impact of ChatGPT availability on the number of git pushes per 100,000 people. The DID estimation shows an increase of approximately 899.268 pushes, which is highly significant with a standard error of 147.395. Both the SC and SDID methods also show positive impacts, with coefficients of 595.059 and 645.623 respectively, though the SC estimate is not statistically significant. The baseline mean outcome is 741.5 pushes per 100,000, suggesting that the introduction of ChatGPT has significantly boosted developer activity in terms of code submissions.

Panel B examines the effect of ChatGPT on the number of repositories created per 100,000 people. The DID estimate shows a substantial and statistically significant increase of 1657.141 repositories, indicating a robust positive effect. However, the SC and SDID estimates, while still positive (464.735 and 204.734 respectively), are not statistically significant. The baseline mean outcome for this measure is 1097.8 repositories per 100,000 people, suggesting that while ChatGPT has a pronounced impact on repository creation as per the DID analysis, the other methods do not confirm the same level of significance.

Panel C focuses on the number of unique developers per 100,000 people. The DID results indicate a significant increase of 578.912 developers, with a standard error of 109.398, highlighting a substantial positive impact. In contrast, the SC and SDID methods show much smaller effects (13.752 and 51.693 respectively), with neither reaching statistical significance. The baseline mean outcome is 527.4 developers per 100,000, suggesting that the presence of ChatGPT has potentially encouraged more individuals to engage in development activities, though the robustness of this finding varies by method.

Overall, the results suggest that the introduction of ChatGPT has generally positive effects on software development activities, particularly in increasing the number of git pushes and repositories created, as evidenced by the DID estimates. However, the magnitude and statistical significance of these effects vary across different methodological approaches. These findings imply that while ChatGPT has likely enhanced developer productivity and engagement, further analysis may be necessary to fully understand the extent and consistency of its impact across different contexts.

We also conducted a similar analysis at the country-programming language level, examining the number of unique developers in each economy who made at least one git push to a repository with a given programming language. From Tables \ref{tab:gpt_high} to \ref{tab:gpt_dsl}, we observe that the introduction of ChatGPT has had a generally positive impact on developer engagement across various programming languages. High-level, general-purpose languages like Python and JavaScript, as well as shell scripting languages, show significant increases in the number of unique developers. Low-level and systems programming languages also benefit, though to a lesser extent. The impact on domain-specific languages varies, with some showing significant positive effects while others remain unaffected or slightly negatively impacted. These findings suggest that ChatGPT has enhanced developer efficiency and engagement, particularly in languages where it can provide substantial assistance with coding tasks.
\section{Discussion}

The rapid progress of Artificial General Intelligence (AGI) is exemplified by the availability of new tools that augment human cognition, enhancing the agility with which certain cognitive tasks can be performed. One area where this utility has become widespread is software development, where large language models (LLMs) are capable of assisting in various stages of the software design and engineering process. This assistance translates into higher productivity, which is crucial for economic growth and prosperity.

Our study found that economies permitting the use of ChatGPT have experienced a significant increase in software development productivity. From a policy perspective, these findings suggest that deregulating access to AI tools could lead to higher productivity and, consequently, foster economic growth. However, there are potential concerns regarding the quality of code generated by ChatGPT, as it is trained on diverse sources, some of which may not contain high-quality code. This could result in increased code churn, potentially offsetting the benefits of faster code creation.

Moreover, the broader social implications must be considered. While current evidence primarily highlights productivity gains through collaborative code writing, future research should also explore the effects of AI availability through other channels. This includes examining the long-term impacts on code quality, developer skills, and the broader economic and social implications of widespread AI tool usage.

In summary, while the integration of AI tools like ChatGPT into software development offers promising productivity benefits, it is essential to balance these gains with considerations of code quality and broader societal impacts. Policymakers and researchers should continue to investigate these dimensions to ensure the sustainable and beneficial integration of AI technologies into the economy.
\newpage

\section{Tables}

\begin{table}[H]
    \caption{Impact of ChatGPT in software development}
    \label{tab:gpt_push_repo_dev}
    \centering
    \scalebox{1}{{
\def\sym#1{\ifmmode^{#1}\else\(^{#1}\)\fi}
\begin{tabular}{l*{3}{c}}
\toprule
                    &\multicolumn{1}{c}{\shortstack{DID}}&\multicolumn{1}{c}{\shortstack{SC}}&\multicolumn{1}{c}{\shortstack{SDID}}\\\cmidrule(lr){2-2}\cmidrule(lr){3-3}\cmidrule(lr){4-4}
\hline
\Gape[0.25cm][0.25cm]{ \underline{Panel A. \textbf{ \textit{Num Pushes per 100k} } } }&               &               &               \\
Chat GPT Available  &     899.268***&     595.059   &     645.623***\\
                    &   (147.395)   &   (437.061)   &   (146.445)   \\

Observations        &        2352   &        2352   &        2352   \\
Baseline Mean Outcome&       741.5   &       741.5   &       741.5   \\

\hline
\Gape[0.25cm][0.25cm]{ \underline{Panel B. \textbf{ \textit{Num Repos per 100k} } } }&               &               &               \\
Chat GPT Available  &    1657.141***&     464.735   &     204.734   \\
                    &   (342.453)   &   (620.581)   &   (133.856)   \\

Observations        &        2352   &        2352   &        2352   \\
Baseline Mean Outcome&      1097.8   &      1097.8   &      1097.8   \\

\hline
\Gape[0.25cm][0.25cm]{ \underline{Panel C. \textbf{ \textit{Num developers per 100k} } } }&               &               &               \\
Chat GPT Available  &     578.912***&      13.752   &      51.693   \\
                    &   (109.398)   &   (204.405)   &    (45.918)   \\

Observations        &        2352   &        2352   &        2352   \\
Baseline Mean Outcome&       527.4   &       527.4   &       527.4   \\
\bottomrule
\end{tabular}
}
}
    \begin{minipage}{13cm}~\\
    \footnotesize Robust standard errors. 
    * $p<$.10, ** $p<$.05, *** $p<$.01.
    Source: Data is created from structured data files of public activity on GitHub, aggregated by economy on a quarterly basis from 2020 onward. Panel A outcome is The number of times developers in a given economy uploaded code to GitHub per 100,000 people. This metric includes all instances where developers pushed code changes to repositories, whether through the command line or GitHub's online platform, with each push potentially containing multiple commits. Panel B outcome is The number of software projects (repositories) in a given economy per 100,000 people. This metric is based on the modal location of all repository members with triage and above access. It includes all repositories, regardless of whether they are actively developed or maintained.Panel C outcome is The number of developer accounts located in a given economy per 100,000 people, based on the modal daily location. This metric excludes bot accounts and users flagged as “spammy” within internal systems. It includes developer accounts that may no longer be active. 
    \end{minipage}
\end{table}
\newpage 

\begin{table}[H]
    \caption{Impact of ChatGPT on Software Development for High-Level, General-Purpose Languages}
    \label{tab:gpt_high}
    \centering
    \scalebox{1}{{
\def\sym#1{\ifmmode^{#1}\else\(^{#1}\)\fi}
\begin{tabular}{l*{3}{c}}
\toprule
                    &\multicolumn{1}{c}{\shortstack{DID}}&\multicolumn{1}{c}{\shortstack{SC}}&\multicolumn{1}{c}{\shortstack{SDID}}\\\cmidrule(lr){2-2}\cmidrule(lr){3-3}\cmidrule(lr){4-4}
\hline
\Gape[0.25cm][0.25cm]{ \underline{Panel A. \textbf{ \textit{Python} } } }&               &               &               \\
Chat GPT Available  &       8.754***&       1.586   &       5.231***\\
                    &     (1.636)   &     (5.387)   &     (1.351)   \\

Observations        &        2256   &        2256   &        2256   \\
Baseline Mean Outcome&       6.137   &       6.137   &       6.137   \\

\hline
\Gape[0.25cm][0.25cm]{ \underline{Panel B. \textbf{ \textit{JavaScript} } } }&               &               &               \\
Chat GPT Available  &      13.568***&       9.484** &       8.257***\\
                    &     (2.536)   &     (4.735)   &     (2.776)   \\

Observations        &        2560   &        2560   &        2560   \\
Baseline Mean Outcome&       12.65   &       12.65   &       12.65   \\

\hline
\Gape[0.25cm][0.25cm]{ \underline{Panel C. \textbf{ \textit{Ruby} } } }&               &               &               \\
Chat GPT Available  &       0.631   &       2.460** &       1.019***\\
                    &     (0.525)   &     (1.234)   &     (0.287)   \\

Observations        &        1840   &        1840   &        1840   \\
Baseline Mean Outcome&       2.831   &       2.831   &       2.831   \\

\hline
\Gape[0.25cm][0.25cm]{ \underline{Panel D. \textbf{ \textit{PHP} } } }&               &               &               \\
Chat GPT Available  &       1.060** &       1.285   &       1.045** \\
                    &     (0.481)   &     (1.363)   &     (0.497)   \\

Observations        &        2128   &        2128   &        2128   \\
Baseline Mean Outcome&       2.577   &       2.577   &       2.577   \\

\hline
\Gape[0.25cm][0.25cm]{ \underline{Panel E. \textbf{ \textit{TypeScript} } } }&               &               &               \\
Chat GPT Available  &       8.421***&       3.887   &       3.863***\\
                    &     (1.591)   &     (2.776)   &     (0.982)   \\

Observations        &        2080   &        2080   &        2080   \\
Baseline Mean Outcome&       2.710   &       2.710   &       2.710   \\
\bottomrule
\end{tabular}
}
}
    \begin{minipage}{10cm}~\\
    \footnotesize Robust standard errors. 
    * $p<$.10, ** $p<$.05, *** $p<$.01.
    Source: Data is created from structured data files of public activity on GitHub, aggregated by economy-programming langugage on a quarterly basis from 2020 onward. -All panels measure the number of unique developers in each economy who made at least one git push to a repository with a given programming language. This measure includes all developers who contributed code, regardless of the size or number of contributions, as long as they made at least one push to a repository.
    \end{minipage}
\end{table}

\begin{table}[H]
    \caption{Impact of ChatGPT on Software Development for Low-Level and Systems Programming Languages}
    \label{tab:gpt_low}
    \centering
    \scalebox{1}{{
\def\sym#1{\ifmmode^{#1}\else\(^{#1}\)\fi}
\begin{tabular}{l*{3}{c}}
\toprule
                    &\multicolumn{1}{c}{\shortstack{DID}}&\multicolumn{1}{c}{\shortstack{SC}}&\multicolumn{1}{c}{\shortstack{SDID}}\\\cmidrule(lr){2-2}\cmidrule(lr){3-3}\cmidrule(lr){4-4}
\hline
\Gape[0.25cm][0.25cm]{ \underline{Panel A. \textbf{ \textit{C} } } }&               &               &               \\
Chat GPT Available  &       3.005***&       3.469** &       1.690***\\
                    &     (0.794)   &     (1.361)   &     (0.564)   \\

Observations        &        1984   &        1984   &        1984   \\
Baseline Mean Outcome&       2.043   &       2.043   &       2.043   \\

\hline
\Gape[0.25cm][0.25cm]{ \underline{Panel B. \textbf{ \textit{C++} } } }&               &               &               \\
Chat GPT Available  &       2.681***&       1.983   &       1.466***\\
                    &     (0.673)   &     (1.692)   &     (0.521)   \\

Observations        &        1856   &        1856   &        1856   \\
Baseline Mean Outcome&       2.684   &       2.684   &       2.684   \\

\hline
\Gape[0.25cm][0.25cm]{ \underline{Panel C. \textbf{ \textit{Rust} } } }&               &               &               \\
Chat GPT Available  &       2.783***&       3.779***&       1.728***\\
                    &     (0.470)   &     (0.795)   &     (0.553)   \\

Observations        &        1104   &        1104   &        1104   \\
Baseline Mean Outcome&       0.402   &       0.402   &       0.402   \\

\hline
\Gape[0.25cm][0.25cm]{ \underline{Panel D. \textbf{ \textit{Assembly} } } }&               &               &               \\
Chat GPT Available  &       0.877***&       1.447***&       0.643***\\
                    &     (0.226)   &     (0.485)   &     (0.181)   \\

Observations        &        1248   &        1248   &        1248   \\
Baseline Mean Outcome&       0.674   &       0.674   &       0.674   \\
\bottomrule
\end{tabular}
}
}
    \begin{minipage}{10cm}~\\
    \footnotesize Robust standard errors. 
    * $p<$.10, ** $p<$.05, *** $p<$.01.
    Source: Data is created from structured data files of public activity on GitHub, aggregated by economy-programming langugage on a quarterly basis from 2020 onward. -All panels measure the number of unique developers in each economy who made at least one git push to a repository with a given programming language. This measure includes all developers who contributed code, regardless of the size or number of contributions, as long as they made at least one push to a repository.
    \end{minipage}
\end{table}

\begin{table}[H]
    \caption{Impact of ChatGPT on Software Development for Shell Scripting Languages}
    \label{tab:gpt_shell}
    \centering
    \scalebox{1}{{
\def\sym#1{\ifmmode^{#1}\else\(^{#1}\)\fi}
\begin{tabular}{l*{3}{c}}
\toprule
                    &\multicolumn{1}{c}{\shortstack{DID}}&\multicolumn{1}{c}{\shortstack{SC}}&\multicolumn{1}{c}{\shortstack{SDID}}\\\cmidrule(lr){2-2}\cmidrule(lr){3-3}\cmidrule(lr){4-4}
\hline
\Gape[0.25cm][0.25cm]{ \underline{Panel A. \textbf{ \textit{Shell} } } }&               &               &               \\
Chat GPT Available  &       5.888***&       8.417** &       4.974***\\
                    &     (1.410)   &     (3.766)   &     (1.506)   \\

Observations        &        2224   &        2224   &        2224   \\
Baseline Mean Outcome&       5.664   &       5.664   &       5.664   \\

\hline
\Gape[0.25cm][0.25cm]{ \underline{Panel B. \textbf{ \textit{Batchfile} } } }&               &               &               \\
Chat GPT Available  &       1.776***&       1.766   &       1.218***\\
                    &     (0.373)   &     (1.554)   &     (0.254)   \\

Observations        &        1584   &        1584   &        1584   \\
Baseline Mean Outcome&       1.752   &       1.752   &       1.752   \\

\hline
\Gape[0.25cm][0.25cm]{ \underline{Panel C. \textbf{ \textit{PowerShell} } } }&               &               &               \\
Chat GPT Available  &       1.262***&       1.719** &       0.841***\\
                    &     (0.228)   &     (0.815)   &     (0.184)   \\

Observations        &        1328   &        1328   &        1328   \\
Baseline Mean Outcome&       0.975   &       0.975   &       0.975   \\
\bottomrule
\end{tabular}
}
}
    \begin{minipage}{10cm}~\\
    \footnotesize Robust standard errors. 
    * $p<$.10, ** $p<$.05, *** $p<$.01.
    Source: Data is created from structured data files of public activity on GitHub, aggregated by economy-programming langugage on a quarterly basis from 2020 onward. -All panels measure the number of unique developers in each economy who made at least one git push to a repository with a given programming language. This measure includes all developers who contributed code, regardless of the size or number of contributions, as long as they made at least one push to a repository.
    \end{minipage}
\end{table}

\begin{table}[H]
    \caption{Impact of ChatGPT on Software Development for Domain-Specific Languages (DSLs)}
    \label{tab:gpt_dsl}
    \centering
    \scalebox{0.85}{{
\def\sym#1{\ifmmode^{#1}\else\(^{#1}\)\fi}
\begin{tabular}{l*{3}{c}}
\toprule
                    &\multicolumn{1}{c}{\shortstack{DID}}&\multicolumn{1}{c}{\shortstack{SC}}&\multicolumn{1}{c}{\shortstack{SDID}}\\\cmidrule(lr){2-2}\cmidrule(lr){3-3}\cmidrule(lr){4-4}
\hline
\Gape[0.25cm][0.25cm]{ \underline{Panel A. \textbf{ \textit{TSQL} } } }&               &               &               \\
Chat GPT Available  &      -0.699***&       0.019   &       0.053   \\
                    &     (0.263)   &     (0.547)   &     (0.400)   \\

Observations        &        1456   &        1456   &        1456   \\
Baseline Mean Outcome&       0.836   &       0.836   &       0.836   \\

\hline
\Gape[0.25cm][0.25cm]{ \underline{Panel B. \textbf{ \textit{PLpgSQL} } } }&               &               &               \\
Chat GPT Available  &       0.344** &       0.564** &       0.270***\\
                    &     (0.144)   &     (0.287)   &     (0.085)   \\

Observations        &         928   &         928   &         928   \\
Baseline Mean Outcome&       0.278   &       0.278   &       0.278   \\

\hline
\Gape[0.25cm][0.25cm]{ \underline{Panel C. \textbf{ \textit{HTML} } } }&               &               &               \\
Chat GPT Available  &      12.347***&       9.674** &       7.981***\\
                    &     (2.347)   &     (4.247)   &     (2.352)   \\

Observations        &        2592   &        2592   &        2592   \\
Baseline Mean Outcome&       15.44   &       15.44   &       15.44   \\

\hline
\Gape[0.25cm][0.25cm]{ \underline{Panel D. \textbf{ \textit{CSS} } } }&               &               &               \\
Chat GPT Available  &      12.303***&       8.378** &       7.775***\\
                    &     (1.753)   &     (4.210)   &     (2.101)   \\

Observations        &        2544   &        2544   &        2544   \\
Baseline Mean Outcome&       11.72   &       11.72   &       11.72   \\

\hline
\Gape[0.25cm][0.25cm]{ \underline{Panel E. \textbf{ \textit{MATLAB} } } }&               &               &               \\
Chat GPT Available  &       0.157** &       1.518***&       0.190*  \\
                    &     (0.074)   &     (0.301)   &     (0.101)   \\

Observations        &         896   &         896   &         896   \\
Baseline Mean Outcome&       0.292   &       0.292   &       0.292   \\

\hline
\Gape[0.25cm][0.25cm]{ \underline{Panel F. \textbf{ \textit{R} } } }&               &               &               \\
Chat GPT Available  &       0.308***&       3.202***&       0.366***\\
                    &     (0.116)   &     (0.598)   &     (0.141)   \\

Observations        &         976   &         976   &         976   \\
Baseline Mean Outcome&       0.251   &       0.251   &       0.251   \\
\bottomrule
\end{tabular}
}
}
    \begin{minipage}{9cm}~\\
    \footnotesize Robust standard errors. 
    * $p<$.10, ** $p<$.05, *** $p<$.01.
    Source: Data is created from structured data files of public activity on GitHub, aggregated by economy-programming langugage on a quarterly basis from 2020 onward. -All panels measure the number of unique developers in each economy who made at least one git push to a repository with a given programming language. This measure includes all developers who contributed code, regardless of the size or number of contributions, as long as they made at least one push to a repository.
    \end{minipage}
\end{table}

\newpage

\section{Figures}

\begin{figure}[H]
    \centering
    \caption{Estimated Trends and Weights for Pushes per 100k}
    \begin{subfigure}{0.45\linewidth}
        \centering
        \includegraphics[width=\linewidth]{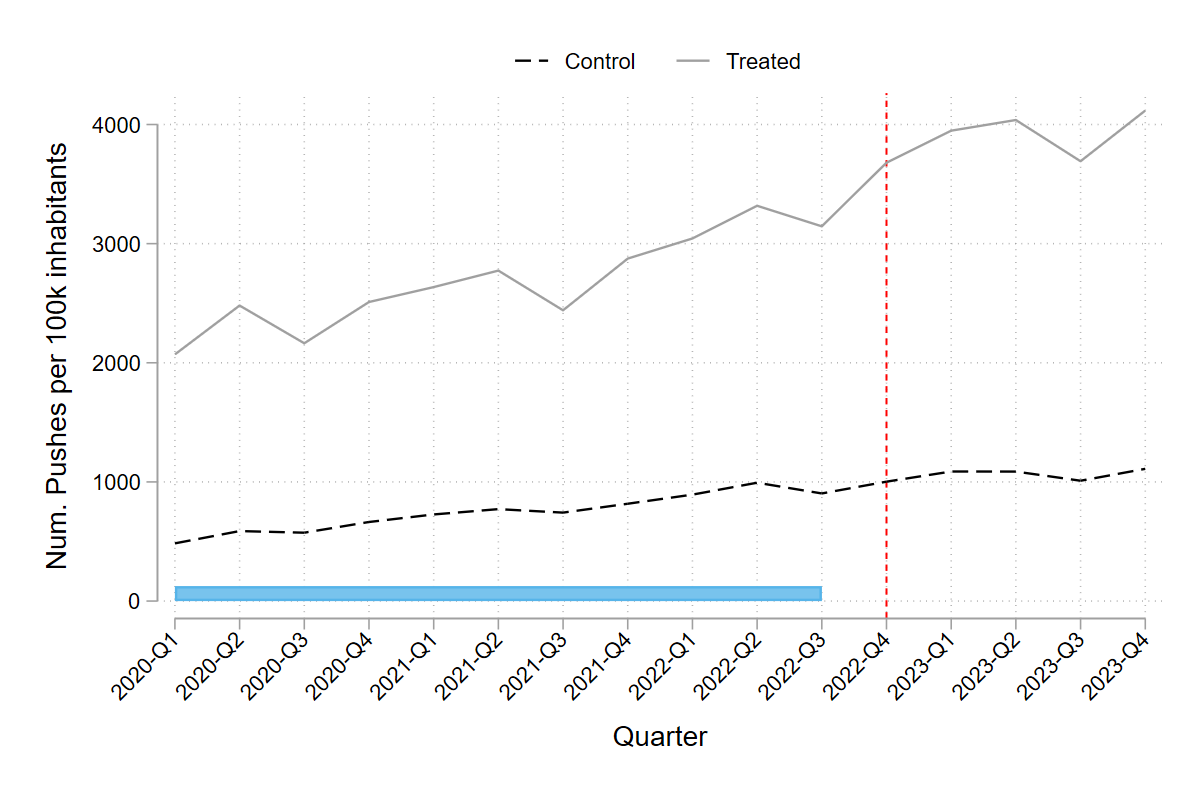}
        \caption{Estimated DiD trends}
        \label{fig:did_trends_pushes}
    \end{subfigure}
    \begin{subfigure}{0.45\linewidth}
        \centering
        \includegraphics[width=\linewidth]{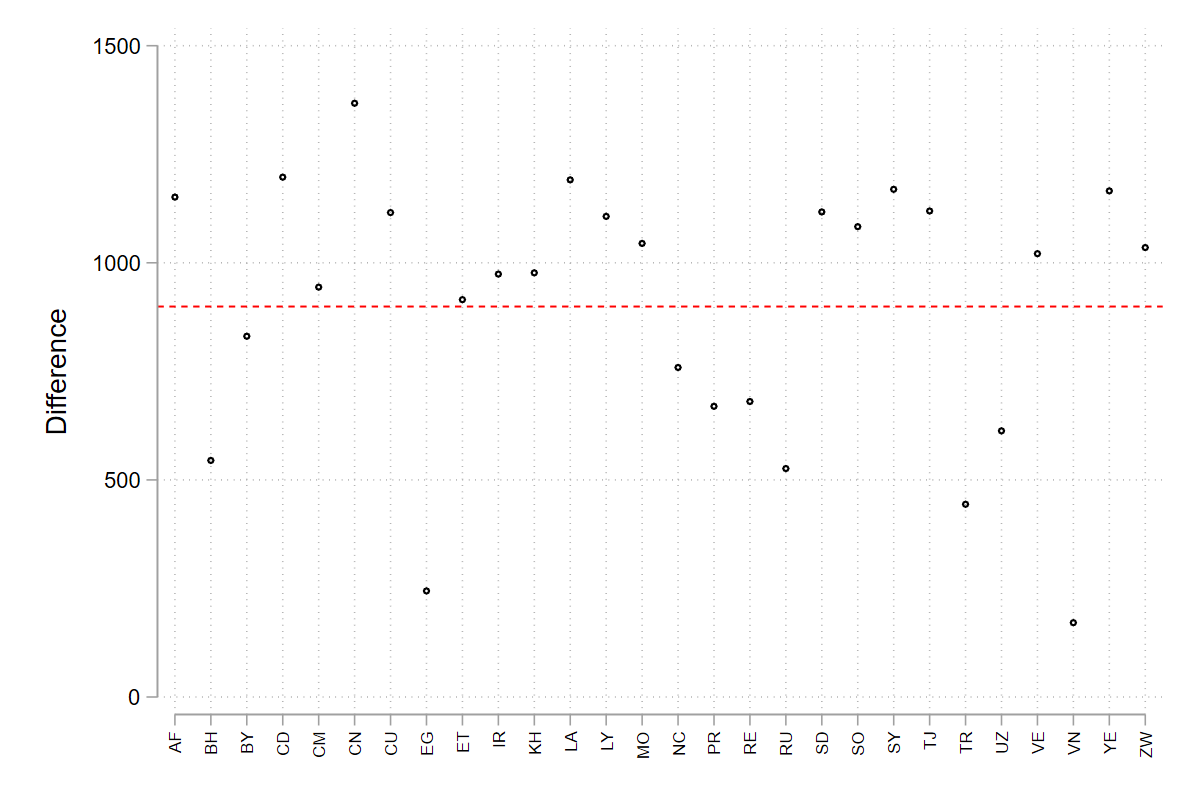}
        \caption{Estimated DiD weights}
        \label{fig:did_weights_pushes}
    \end{subfigure}
    \begin{subfigure}{0.45\linewidth}
        \centering
        \includegraphics[width=\linewidth]{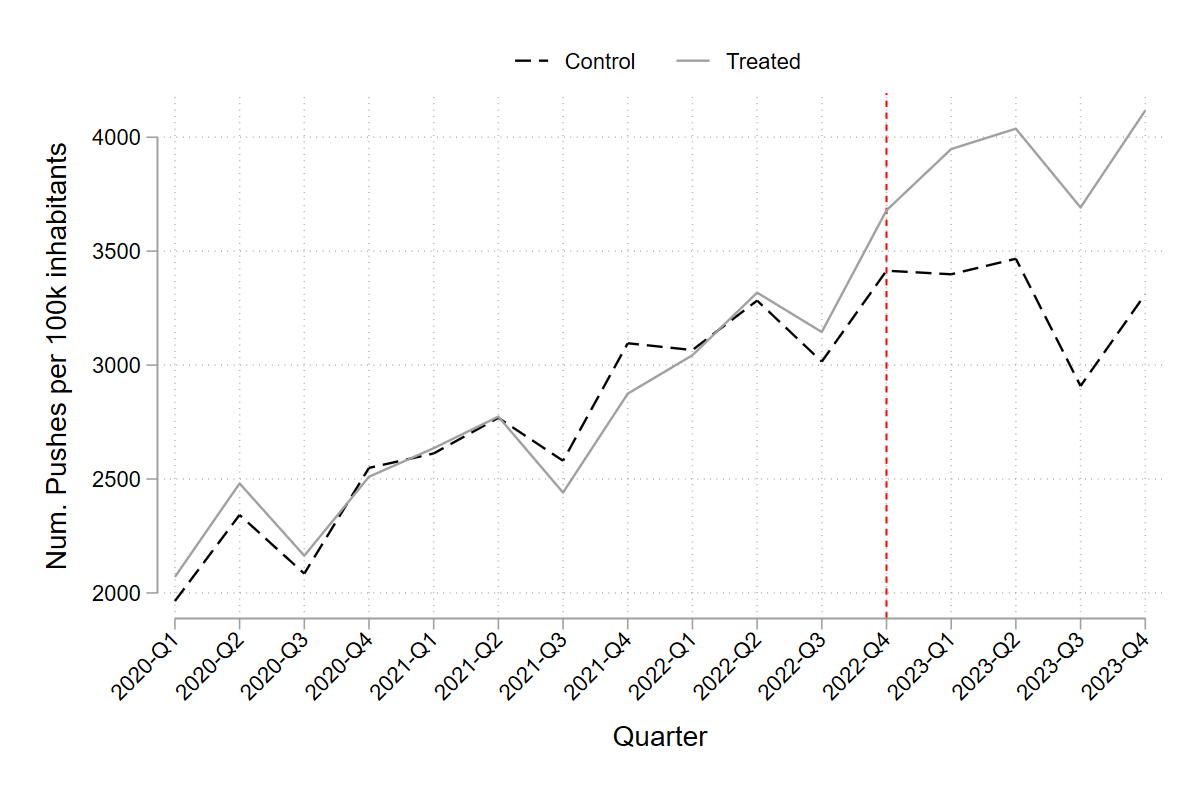}
        \caption{Estimated SC trends}
        \label{fig:sc_trends_pushes}
    \end{subfigure}
    \begin{subfigure}{0.45\linewidth}
        \centering
        \includegraphics[width=\linewidth]{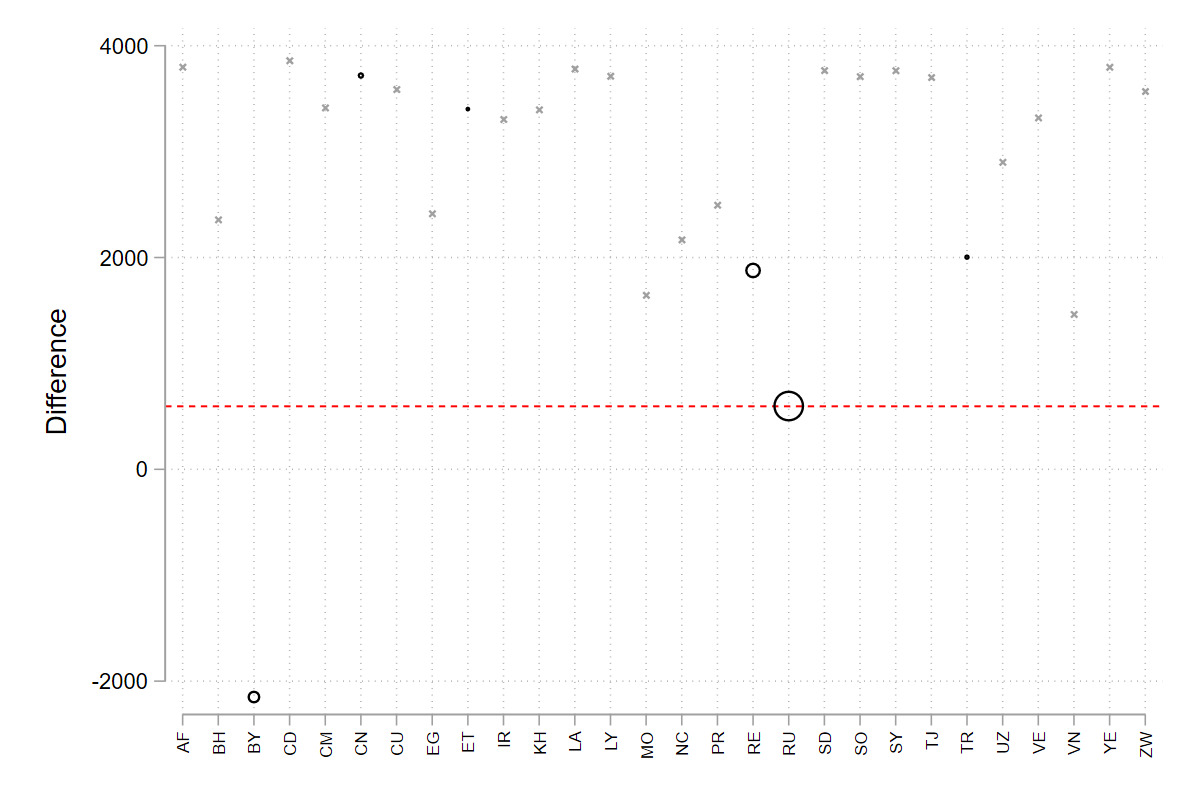}
        \caption{Estimated SC weights}
        \label{fig:sc_weights_pushes}
    \end{subfigure}
    \begin{subfigure}{0.45\linewidth}
        \centering
        \includegraphics[width=\linewidth]{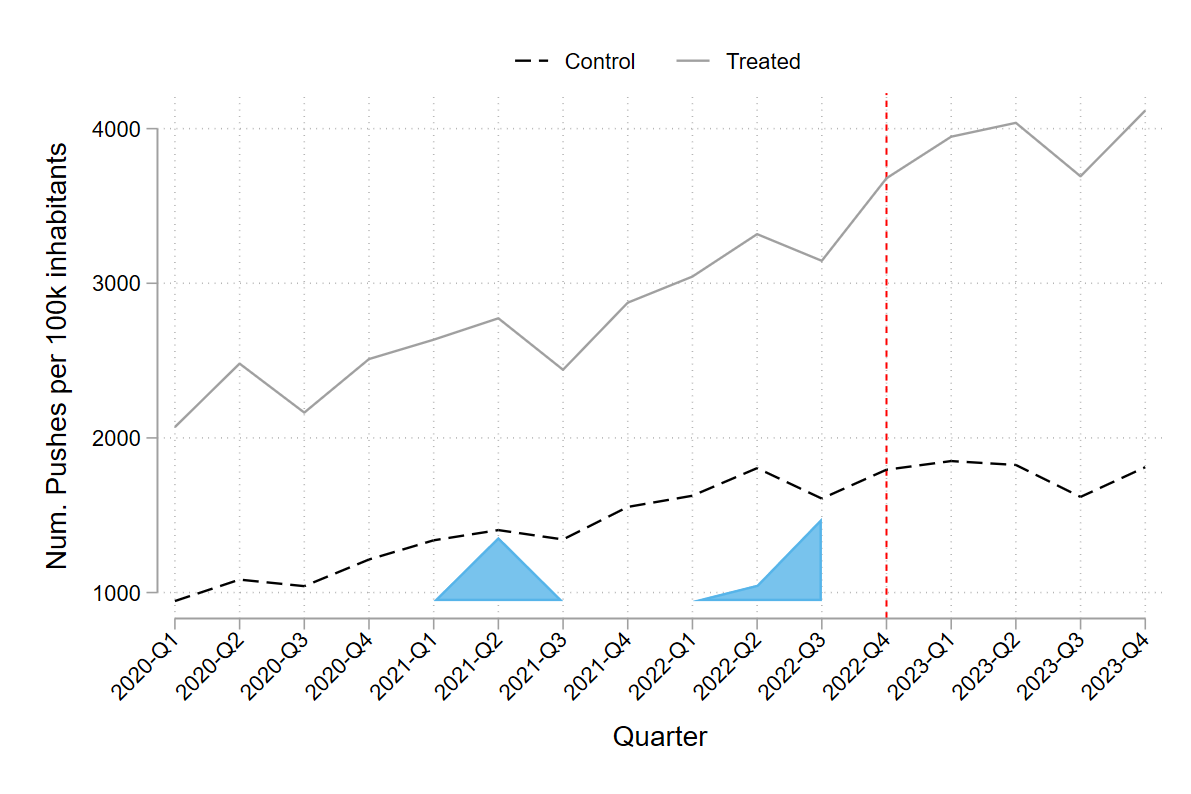}
        \caption{Estimated SDiD trends}
        \label{fig:sdid_trends_pushes}
    \end{subfigure}
    \begin{subfigure}{0.45\linewidth}
        \centering
        \includegraphics[width=\linewidth]{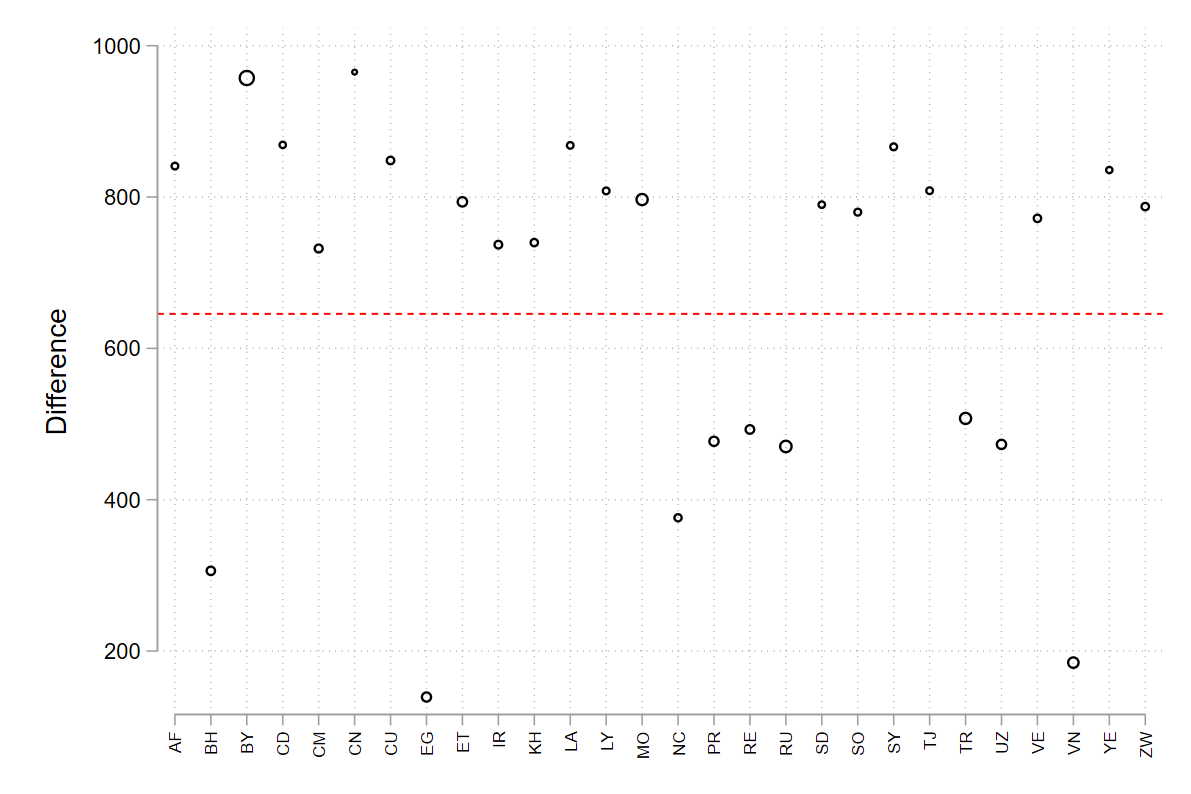}
        \caption{Estimated SDiD weights}
        \label{fig:sdid_weights_pushes}
    \end{subfigure}
    \label{fig:trend_weight_push}
\end{figure}

\begin{figure}[H]
    \centering
    \caption{Estimated Trends and Weights for Repositories per 100k}
    \begin{subfigure}{0.45\linewidth}
        \centering
        \includegraphics[width=\linewidth]{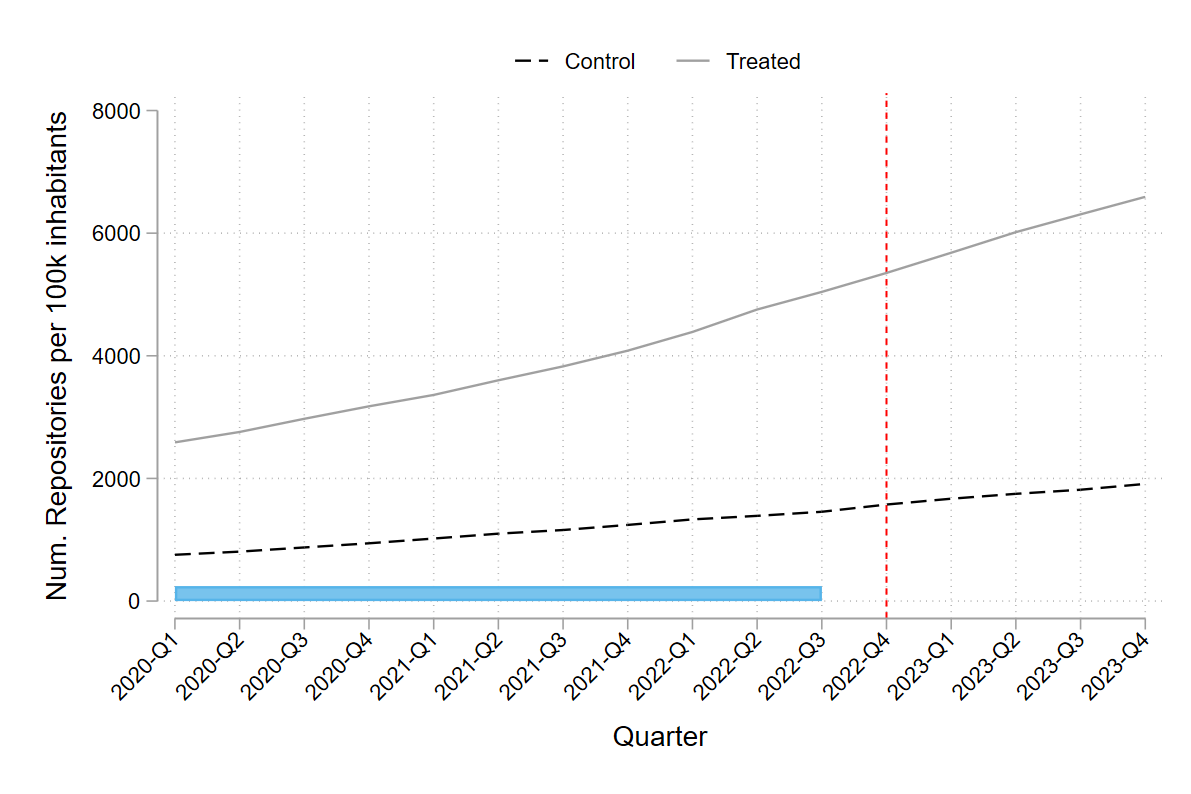}
        \caption{Estimated DiD trends}
        \label{fig:did_trends_repo}
    \end{subfigure}
    \begin{subfigure}{0.45\linewidth}
        \centering
        \includegraphics[width=\linewidth]{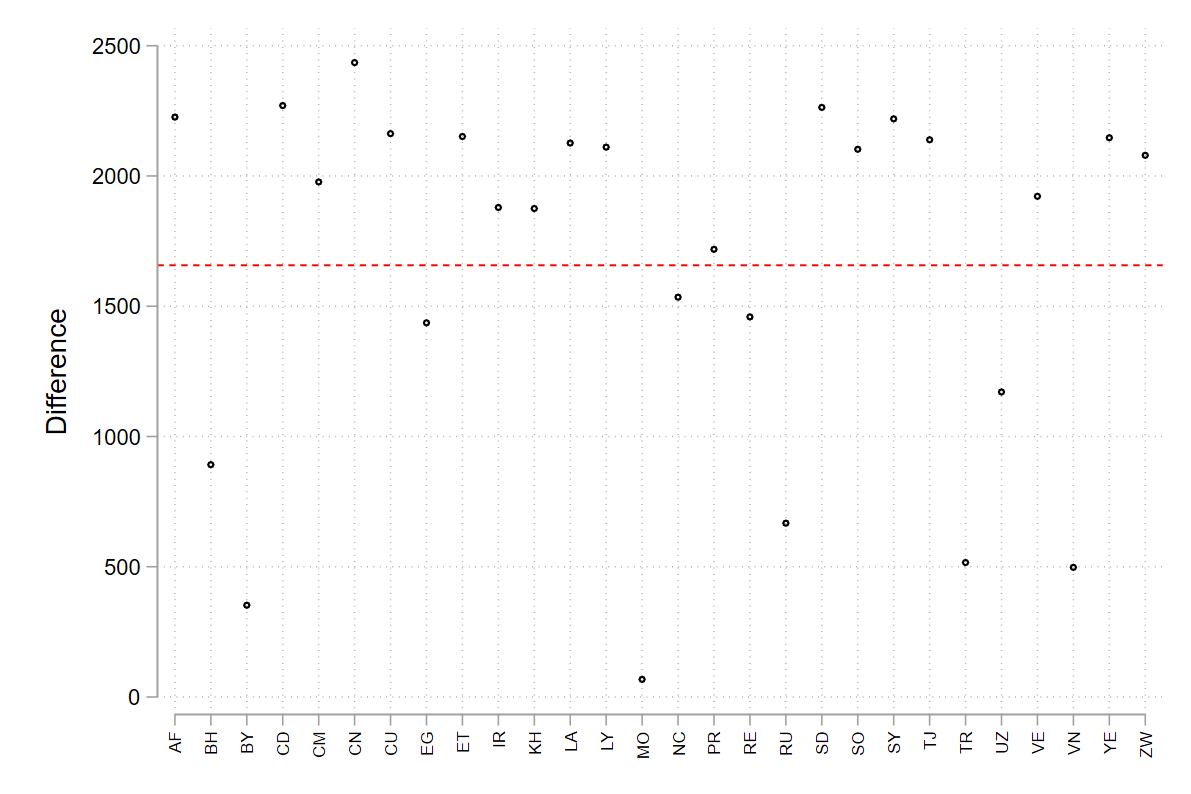}
        \caption{Estimated DiD weights}
        \label{fig:did_weights_repo}
    \end{subfigure}
    \begin{subfigure}{0.45\linewidth}
        \centering
        \includegraphics[width=\linewidth]{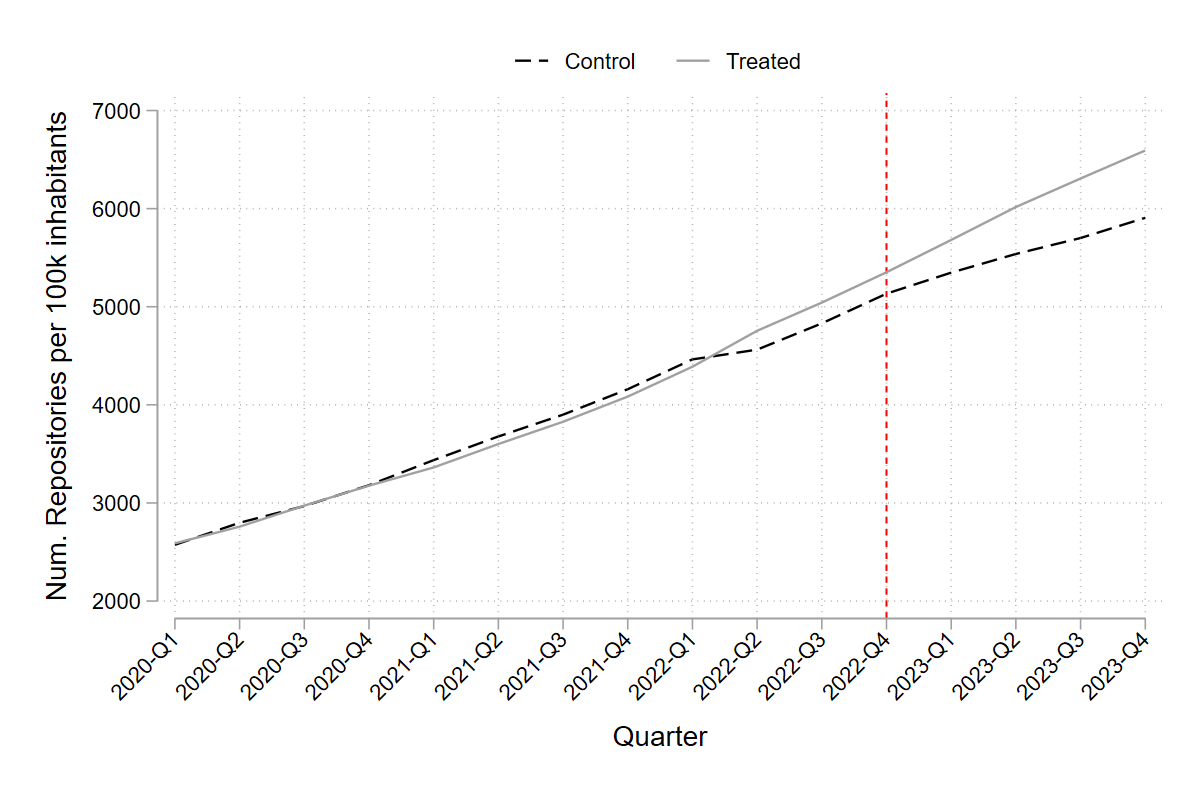}
        \caption{Estimated SC trends}
        \label{fig:sc_trends_repo}
    \end{subfigure}
    \begin{subfigure}{0.45\linewidth}
        \centering
        \includegraphics[width=\linewidth]{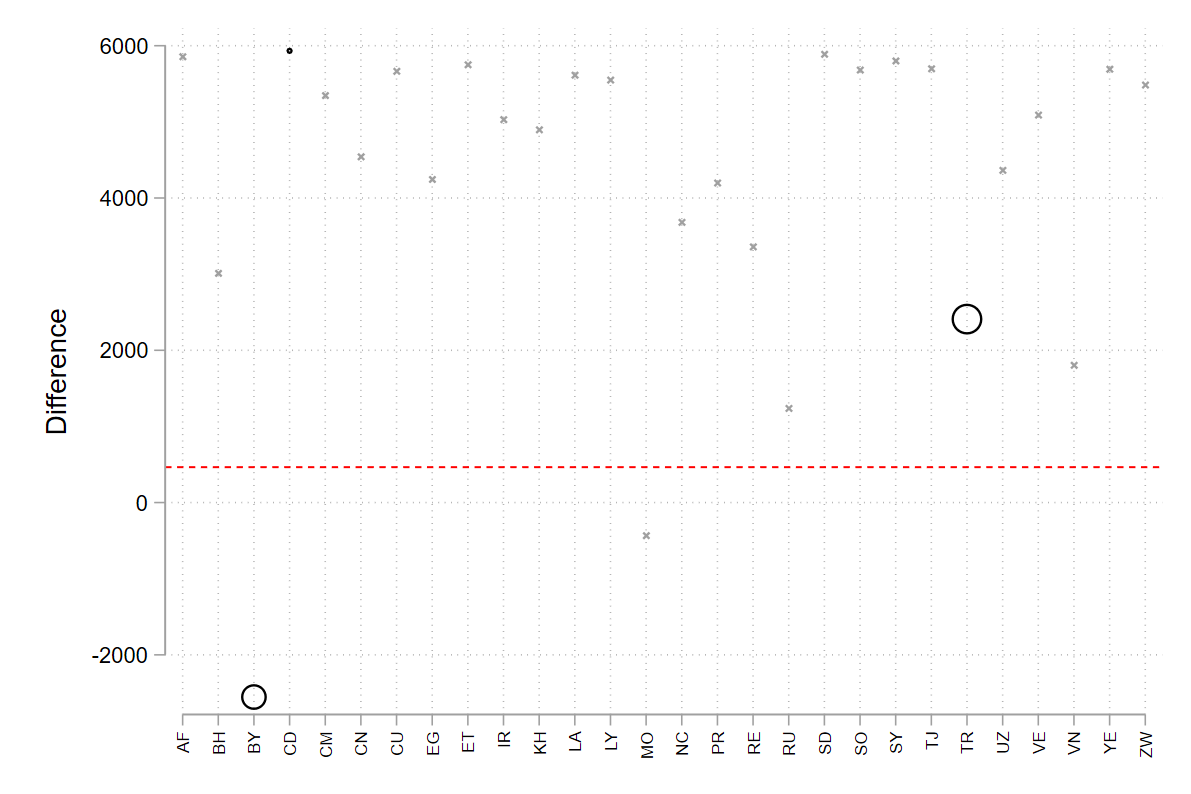}
        \caption{Estimated SC weights}
        \label{fig:sc_weights_repo}
    \end{subfigure}
    \begin{subfigure}{0.45\linewidth}
        \centering
        \includegraphics[width=\linewidth]{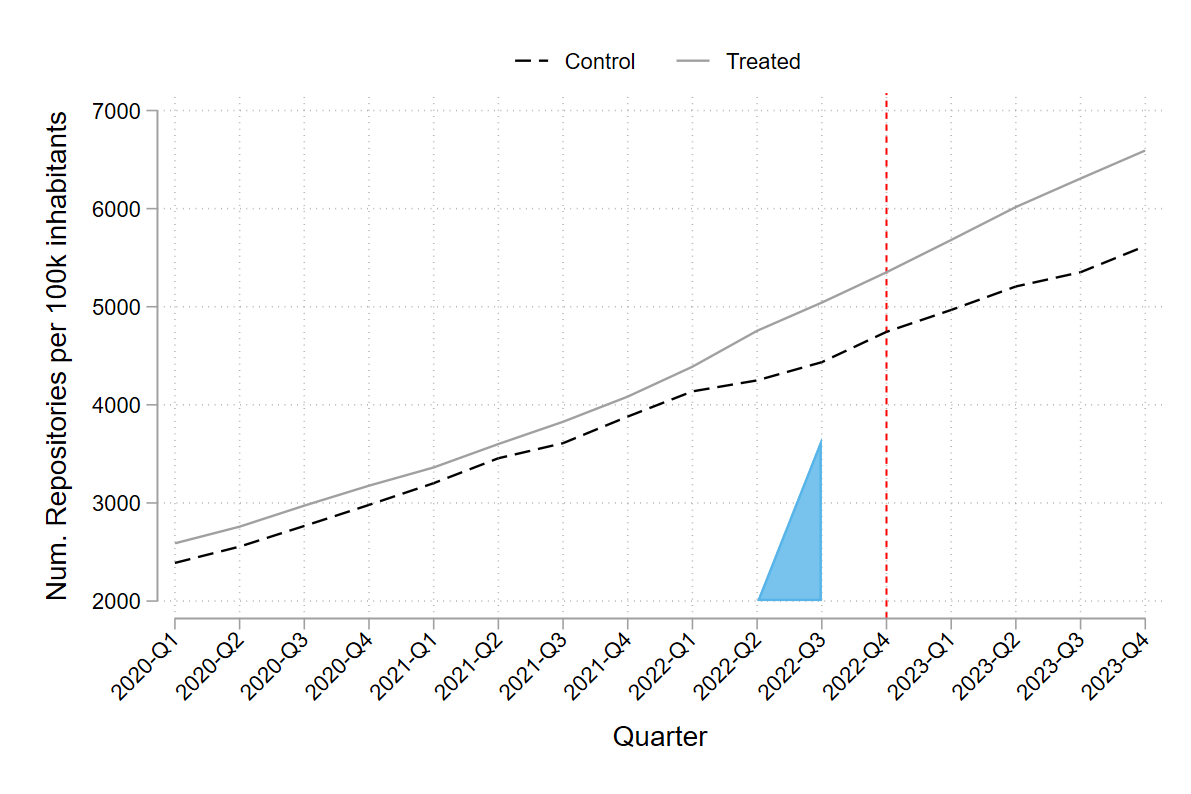}
        \caption{Estimated SDiD trends}
        \label{fig:sdid_trends_repo}
    \end{subfigure}
    \begin{subfigure}{0.45\linewidth}
        \centering
        \includegraphics[width=\linewidth]{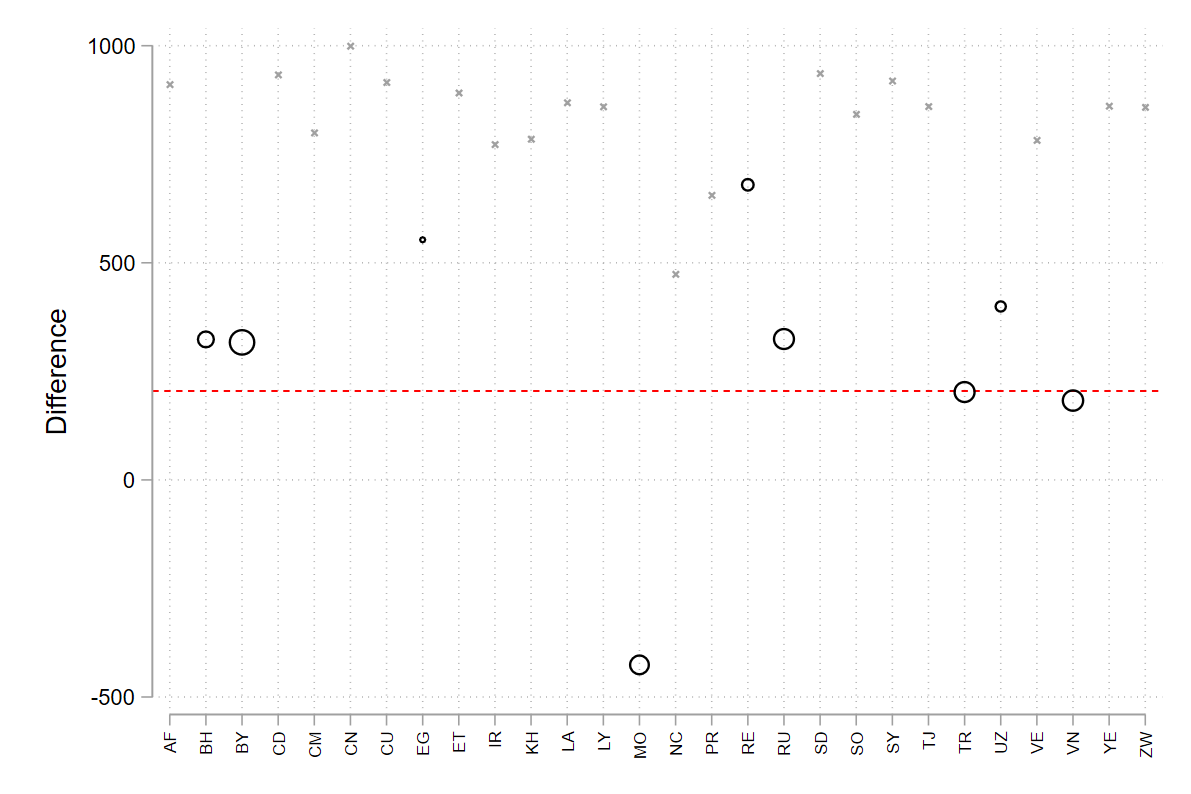}
        \caption{Estimated SDiD weights}
        \label{fig:sdid_weights_repo}
    \end{subfigure}
    \label{fig:trend_weight_repo}
\end{figure}

\begin{figure}[H]
    \centering
    \caption{Estimated Trends and Weights for Developers per 100k}
    \begin{subfigure}{0.45\linewidth}
        \centering
        \includegraphics[width=\linewidth]{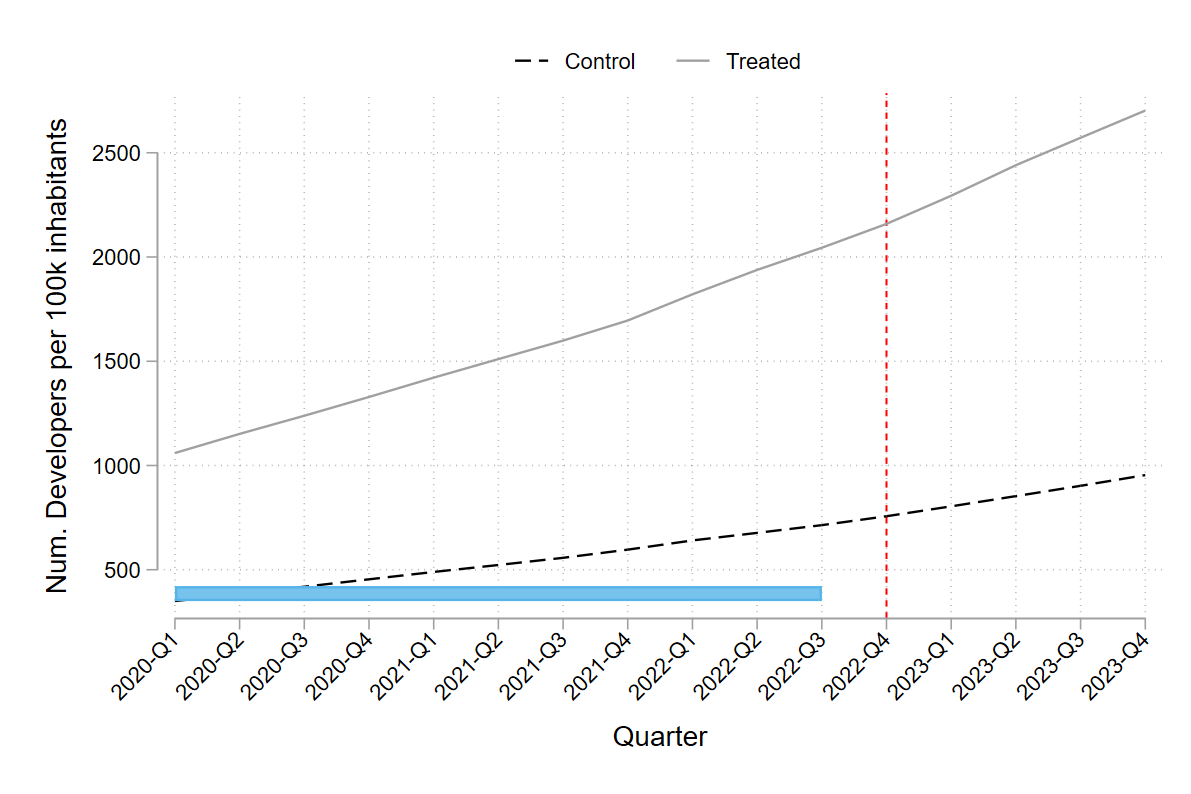}
        \caption{Estimated DiD trends}
        \label{fig:did_trends_dev}
    \end{subfigure}
    \begin{subfigure}{0.45\linewidth}
        \centering
        \includegraphics[width=\linewidth]{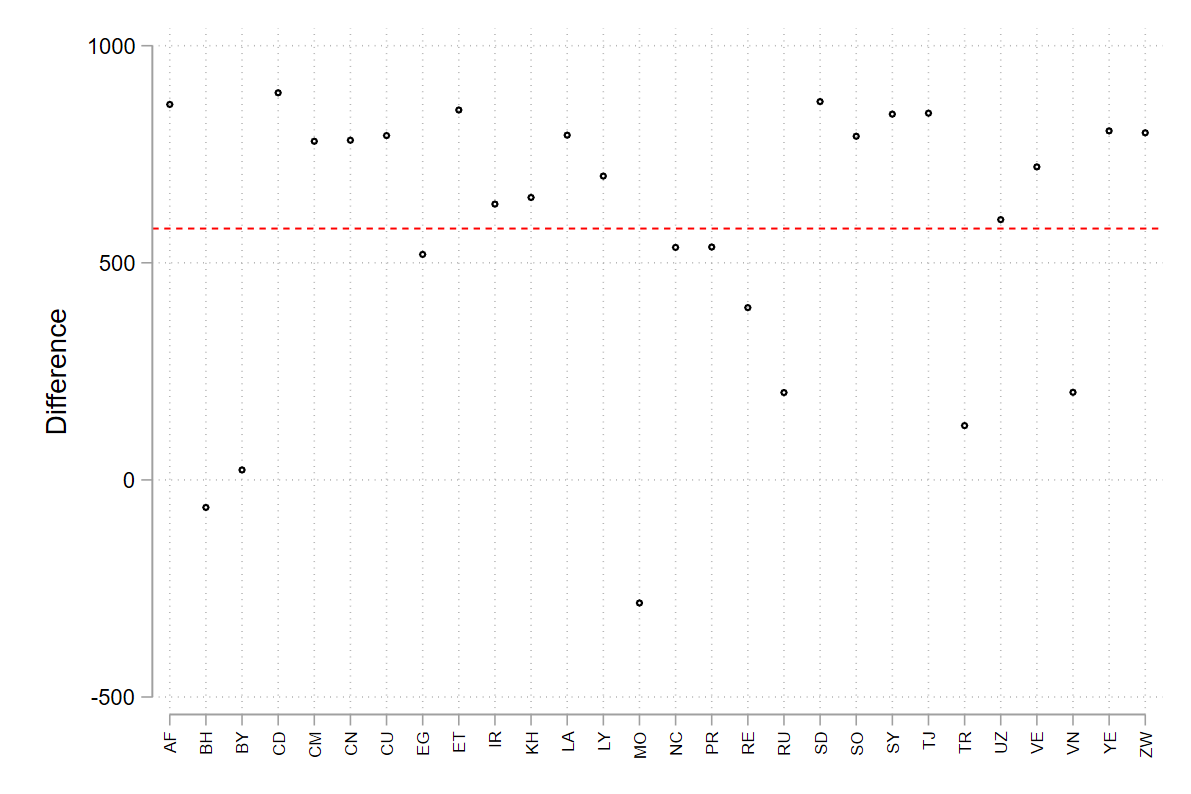}
        \caption{Estimated DiD weights}
        \label{fig:did_weights_dev}
    \end{subfigure}
    \begin{subfigure}{0.45\linewidth}
        \centering
        \includegraphics[width=\linewidth]{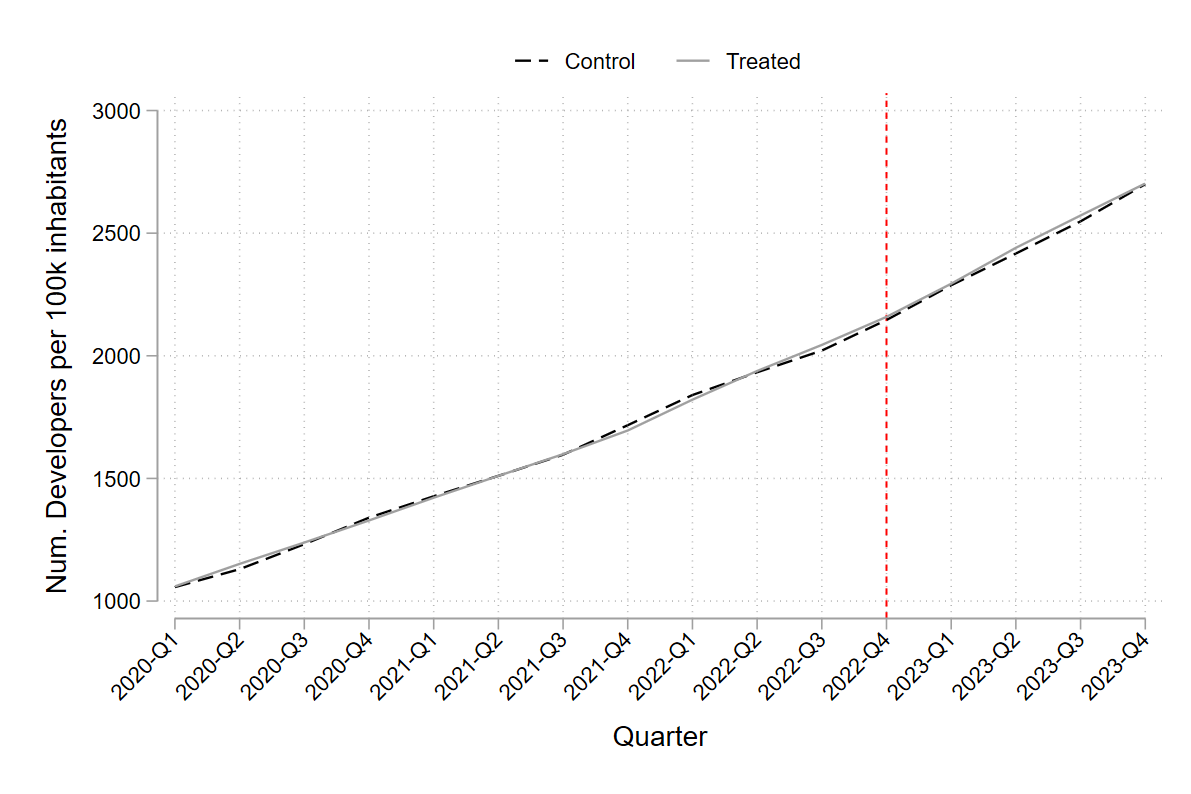}
        \caption{Estimated SC trends}
        \label{fig:sc_trends_dev}
    \end{subfigure}
    \begin{subfigure}{0.45\linewidth}
        \centering
        \includegraphics[width=\linewidth]{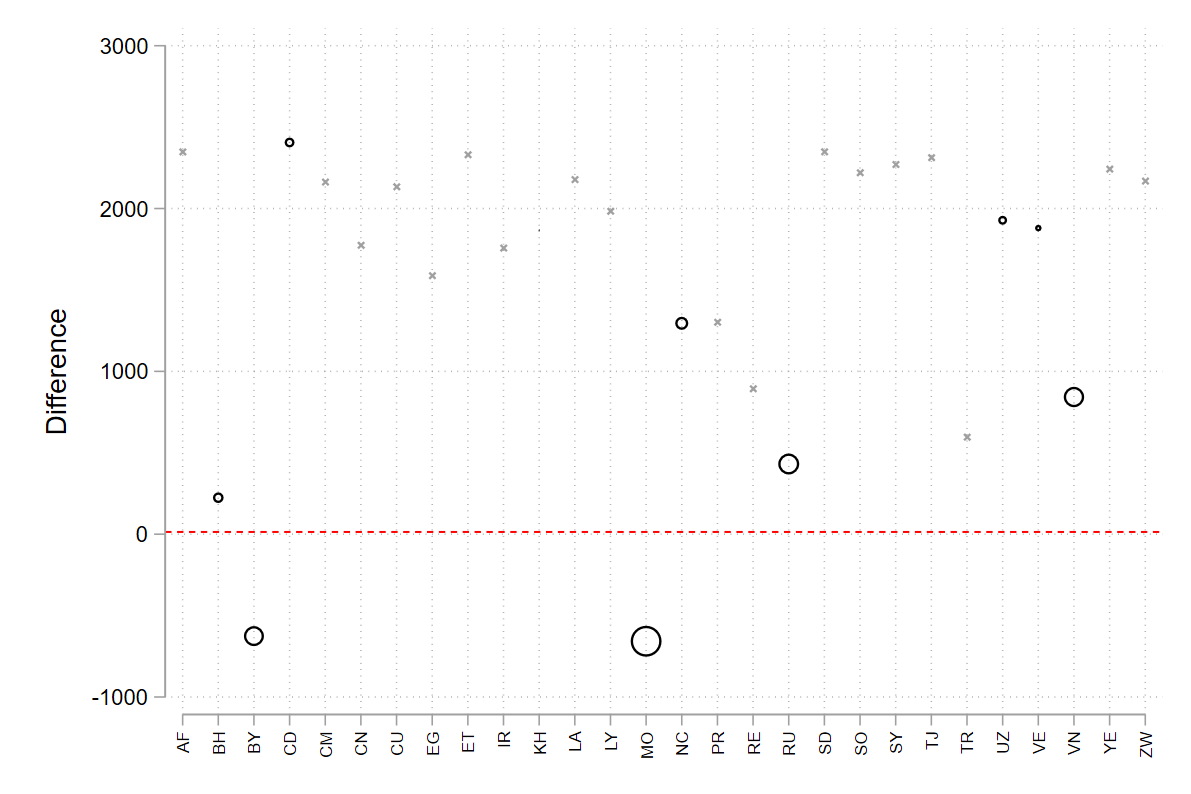}
        \caption{Estimated SC weights}
        \label{fig:sc_weights_dev}
    \end{subfigure}
    \begin{subfigure}{0.45\linewidth}
        \centering
        \includegraphics[width=\linewidth]{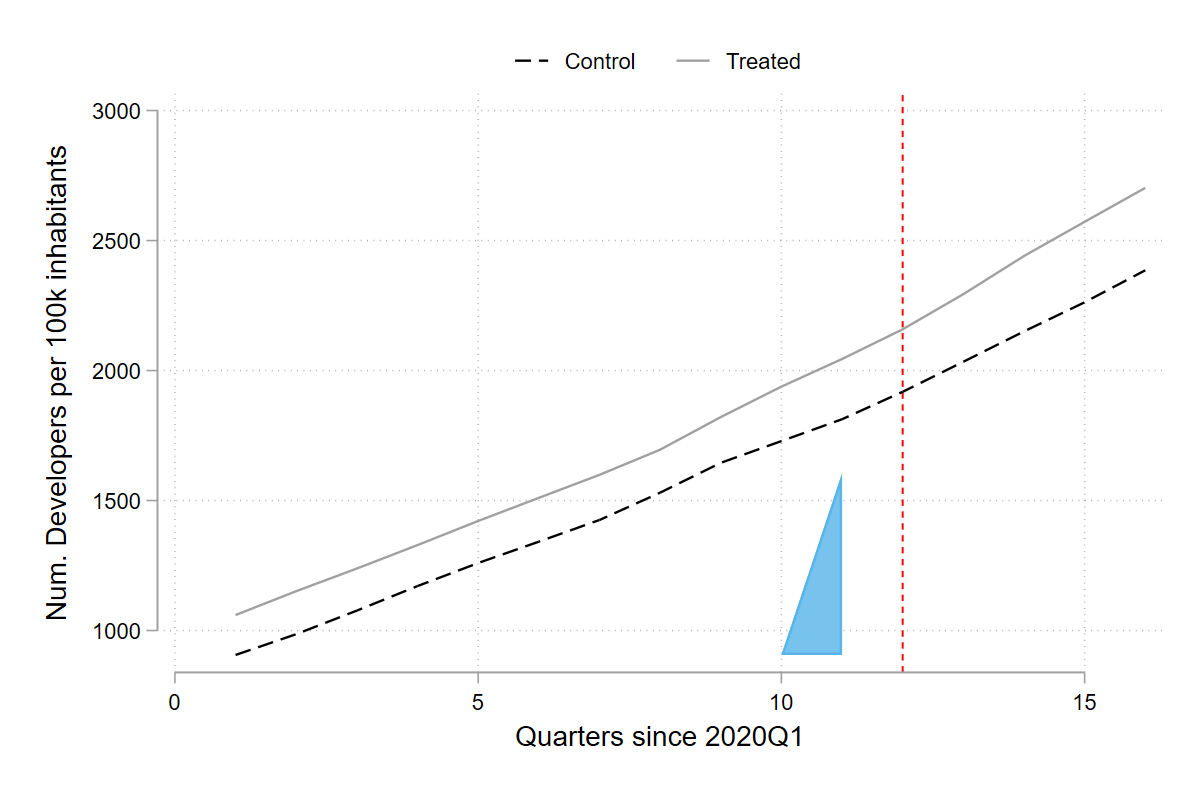}
        \caption{Estimated SDiD trends}
        \label{fig:sdid_trends_dev}
    \end{subfigure}
    \begin{subfigure}{0.45\linewidth}
        \centering
        \includegraphics[width=\linewidth]{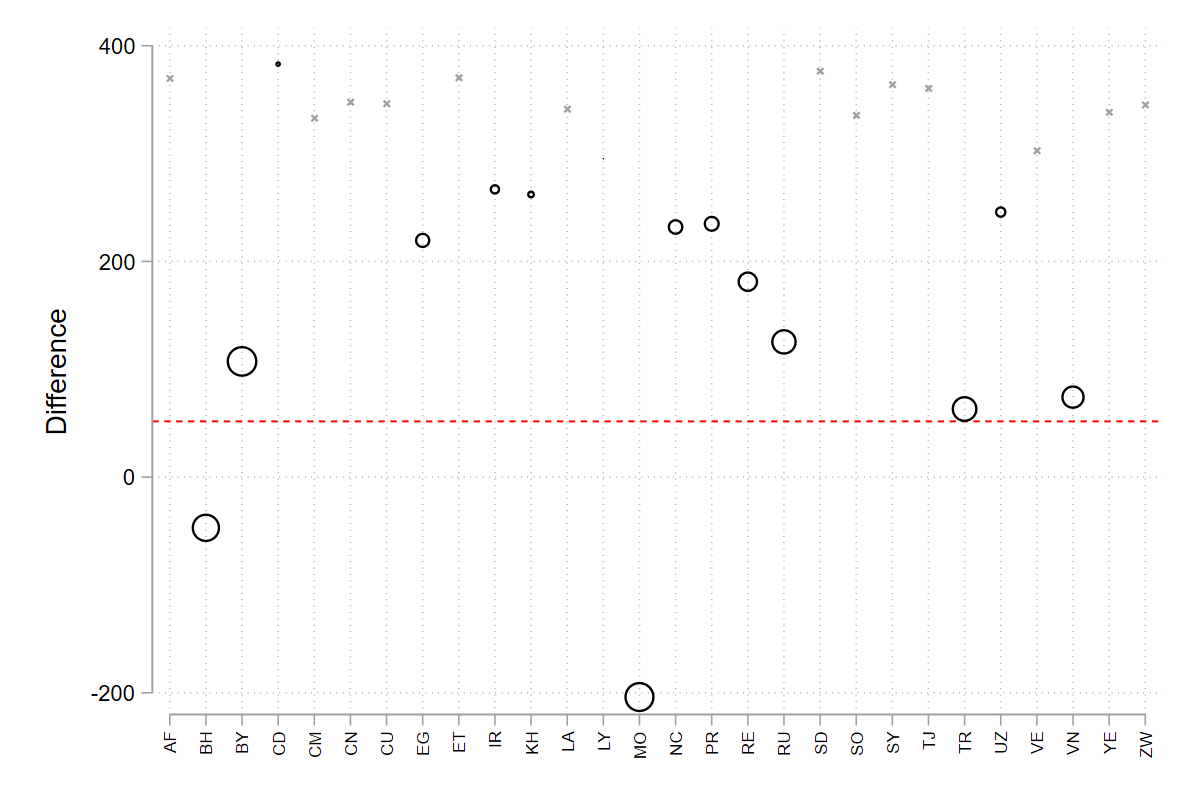}
        \caption{Estimated SDiD weights}
        \label{fig:sdid_weights_dev}
    \end{subfigure}
    \label{fig:trend_weight_dev}
\end{figure}

\newpage
\bibliography{references}

\end{document}